\documentclass{aa}

\usepackage{graphicx}
\usepackage{txfonts}

\begin{document}

\title{Photometry of the SW Sex-type nova-like BH Lyncis in high state\thanks{Based on observations obtained at Rozhen
National Astronomical Observatory, Bulgaria}}
\author{
V. Stanishev\inst{1,2}$^{\star\star}$
\and
Z. Kraicheva\inst{2}$^{\star\star}$
\and
V. Genkov\inst{2}\thanks{E-mail:
{\rm vall,\#zk,\#vgenkov@astro.bas.bg} (VS,\#ZK,\#VG)}
}


\institute{Department of Physics, Stockholm University, Albanova University Center, 106 91 Stockholm, Sweden
\and
Institute of Astronomy, Bulgarian Academy of Sciences,
           72 Tsarighradsko Shousse Blvd., 1784 Sofia, Bulgaria}

\date{Received ; accepted }

\authorrunning{V. Stanishev et al.}

\abstract{}
 {We present a photometric study of the deeply eclipsing 
SW Sex-type nova-like cataclysmic variable star \object{BH Lyn}.}
 {Time-resolved $V$-band CCD photometry was obtained for seven nights between
  1999 and 2004.}
 {We determined 11 new eclipse timings of \object{BH Lyn} and derived 
  a refined orbital ephemeris with an orbital period of 0\fd155875577(14). 
 During the observations,  \object{BH Lyn} 
 was in high-state with $V\simeq15.5$ mag. The star presents  $\sim1.5$ mag deep
 eclipses with mean full-width at half-flux of $0.0683(\pm0.0054)P_{orb}$. 
 The eclipse shape is highly variable, even changing form cycle to cycle.
 This is most likely due to accretion disc surface brightness 
 distribution variations, most probably caused by strong flickering.
 Time-dependent accretion disc self-occultation or variations of the hot spot(s) intensity 
 are also possible explanations. Negative superhumps with period of $\sim0\fd145$ 
 are detected in two long runs in 2000. A possible connection between SW Sex and 
 negative superhump phenomena through the presence of tilted accretion disc 
 is discussed, and a way to observationally test this is suggested.}
 {}


\maketitle

\section{Introduction}

\object{BH Lyncis} is an eclipsing novalike (NL) cataclysmic
variable (CV) with an orbital period of $\sim3\fh74$ (Andronov et.
\cite{andr89}). Thorstensen et al. (\cite{th91a}), Dhillon et al.
(\cite{d92}), and Hoard \& Szkody (\cite{hs}) have shown that
spectral behavior of \object{BH Lyn} resembles that of \object{SW
Sex}-type novalikes. \object{SW Sex} stars are spectroscopically
defined sub-class of novalikes (Thorstensen et al. \cite{th91b}).
Most of them are eclipsing, but show
single-peaked emission lines contrary to the expected
double-peaked from high-inclined accretion discs. Other
distinctive characteristics are high-velocity emission components, 
narrow absorption components superimposed over emission lines around
orbital phase 0.5, and a large phase offset of the emission line
radial velocities, with respect to the photometric conjunction.
The eclipse profiles are V-shaped rather that U-shaped, and 
 the accretion discs brightness temperature
distribution derived from eclipse mapping
 is much flatter than expected for a
steady-state accretion disc (e.g. Rutten et al. \cite{rutt}). 
Patterson (\cite{patt99}) reports that most of the SW Sex stars
show both negative and positive superhumps. Besides, some of
the members show low states (Honeycutt et al. \cite{hon}).
Currently, there is no widely accepted model of \object{SW Sex}
stars. In most of the CVs, the accretions stream from secondary 
hits on the outer disc edge, and a hot spot is formed at the impact. 
In the most elaborated model of the SW Sex stars, 
Hellier (\cite{hel98}) suggested that part of the gas in
the stream does not stop in the vicinity of the hot spot. 
Instead, it continues moving above the disc surface, hits
the disc close to the white dwarf, and thus forms a second spot.
Recently, Rodriguez-Gil et
al. (\cite{rg}) discovered variable circular polarization in
\object{LS Peg} and suggested that \object{SW Sex} stars are
intermediate polars with the highest mass accretion rates.

The object of this study, \object{BH Lyn}, is mostly studied 
spectroscopically, and the existing
photometric data are generally used to obtain the eclipse
ephemeris and to supplement the spectral observations. In this
paper, we report the results of our photometry of \object{BH Lyn}
obtained in 1999-2004.

\section{Observations and data reduction}

\begin{table}
\caption[]{$V$ band observations of \object{BH Lyn}. The 
eclipse timings are also given.}
\begin{tabular}{cccc}
\hline
\hline
\noalign{\smallskip}
 UT date & HJD  Start & Duration  &  HJD mid-eclipse \\
         & -2451000 & [hour]  &  -2451000 \\
\noalign{\smallskip}
\hline
\noalign{\smallskip}
 Feb. 20, 1999 &  230.3833 & 3.23  &   230.45114  \\
 Jan. 08, 2000 &  552.2748 & 5.61  &   552.33385  \\
               &           &       &   552.48983  \\
 Jan. 09, 2000 &  553.2240 & 9.16  &   553.26945  \\
               &           &       &   553.42513  \\
               &           &       &   553.58089  \\
 Mar. 12, 2000 &  616.2509 & 4.21  &   616.39900  \\
 Feb. 28, 2003 & 1699.3052 & 7.44  &  1699.42318  \\
               &           &       &  1699.57916  \\
 Dec. 19, 2003 & 1993.5075 & 3.77  &  1993.56007  \\
 Jan. 18, 2004 & 2023.5165 & 3.86  &  2023.64410  \\
\noalign{\smallskip}
\hline
\end{tabular}
\label{obs}
\end{table}

\begin{figure*}[ht]
\centering
\includegraphics*[width=18cm]{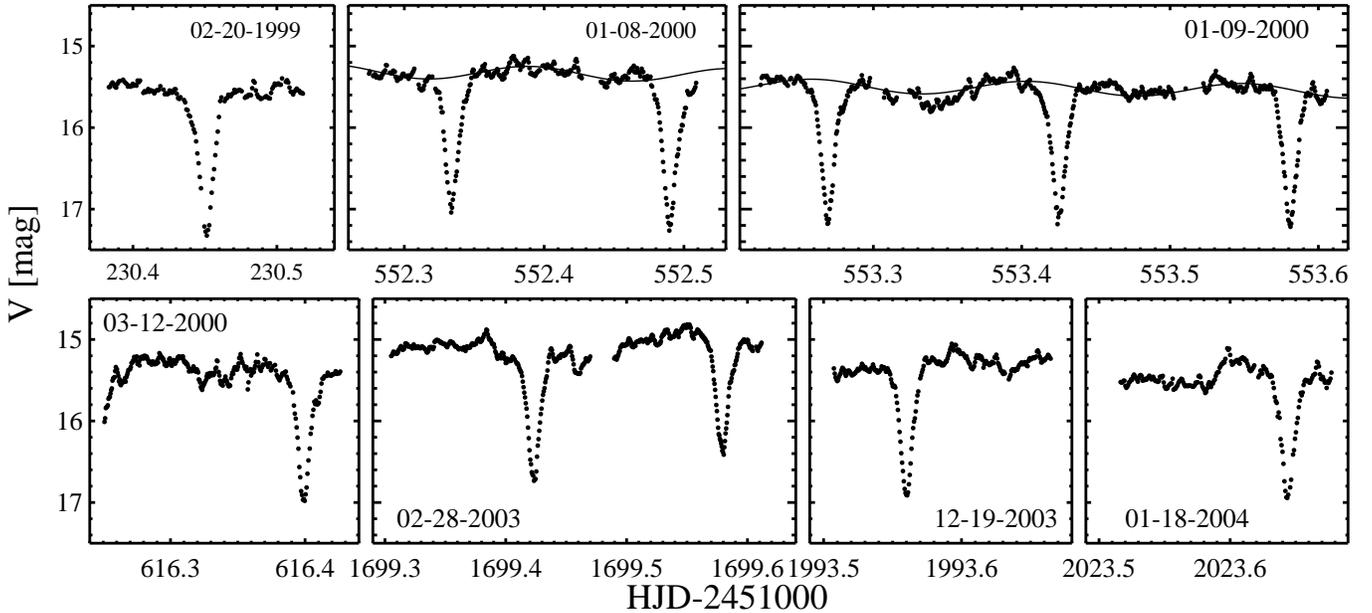}
\caption{$V$-band observations of \object{BH Lyn}. The solid line shows 
the sinusoidal fit with the period of the superhumps detected in the 2000
data.}
\label{data}
\end{figure*}

The photometric observations of \object{BH Lyn} were obtained with
the 2.0-m telescope in the Rozhen Observatory. A Photometrics 1024$^2$
CCD camera and a Johnson $V$ filter were used. The CCD camera was
2$\times$2 pixels binned, which resulted in $\sim$13 s of read-out
dead-time. In total, 7 runs of photometric data were obtained 
between 1999 and 2004. The
exposure time used was between 30 and 60 s. Some details of the
observations are given in Table\,\ref{obs}. After bias and
flat-field corrections, the photometry was done with the standard
DAOPHOT aperture photometry procedures (Stetson \cite{ste}). The
magnitude of \object{BH Lyn} was measured relative to the star
BH Lyn-5 ($V=14.47$), and BH Lyn-4 ($V=15.30$) served as
a check (Henden \& Honeycutt \cite{comp}). The runs are shown in
Fig.\,\ref{data}, and it appears that  \object{BH Lyn} was in high state 
during all observations.

\section{Results}

\begin{figure}[t]
\includegraphics*[width=8.8cm]{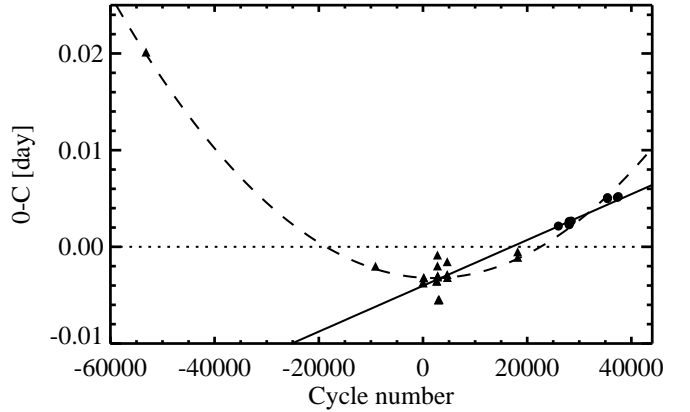}   
\centering
\caption{$O-C$ residuals of the minima with respect to the 
best linear ephemeris. The second-order
polynomial fit to the $O-C$ residuals is also shown. The solid line is our best 
linear ephemeris. The filled circles show our new timings.}
\label{oc}
\end{figure}

The eclipse timings given in Table\,\ref{obs} were determined by
fitting a parabola to the lower half of the eclipses. To refine the
orbital ephemeris of \object{BH Lyn}, we also used the eclipse
timings collected by Hoard \& Szkody (\cite{hs}). The $O-C$
residuals with respect to the best linear ephemeris are shown in
Fig.\,\ref{oc}. Clearly, the linear ephemeris does not
describe the eclipse times  well and, as  
Hoard \& Szkody (\cite{hs}) point out, this is mainly due to the
anomalously large, positive residual of the first eclipse timing.
Hoard \& Szkody (\cite{hs}) suggested that the first eclipse timing 
was in error and calculated a linear
ephemeris without it.
 The $O-C$ residuals of our new eclipse
timings are rather large, 
$\sim$0\fd006, and increasingly positive.
Together with the first two timings, whose $O-C$ residuals are also positive, 
 this suggests the presence of a curvature in the $O-C$ residuals. 
 The dashed line is the second-order polynomial fit to all eclipse timings.
The quadratic term is $7.6\times10^{-12}$ and implies that the orbital period
of \object{BH Lyn} increases on a time scale of
$\sim4.2\times10^6$ yrs. In most of the CVs, the mass donor star 
is the less massive one, and hence, if the mass transfer is conservative,
 the orbital period of the system will increase. 
For plausible component masses in \object{BH
Lyn}, $M_{WD}\sim0.73$ and  $M_2\sim0.33$ (Hoard \& Szkody
\cite{hs}), the mass
transfer rate should be $\dot{M}\sim5\times10^{-8}\,M_{\sun}$\,yr$^{-1}$
 to be compatible with the putative orbital period increase.
However, 
there are several arguments against this scenario.
First, there is a bulk of evidence that CVs evolve toward shorter orbital
periods due to the angular momentum loss of the secondary by  magnetic breaking
(Warner \cite{war}). Second,  $\dot{M}\sim5\times10^{-8}\,M_{\sun}$\,yr$^{-1}$ 
is   probably too high and generally not typical for CVs.
 Third, the eclipse timings presented by 
Andronov et al. (\cite{andr89}) have been determined by the phase 
folding of observations with photographic plates with rather long exposure 
times of 8, 12, and  30 min. It is not surprising then, that 
 those timings exhibit relatively large scatter 
(the timings with cycle numbers $\sim3000$). 
The second timing has been determined from plates with 
exposure time 30 min, only slightly shorter than the total eclipse 
duration, and its  large positive $O-C$ of this timing may be 
a statistical fluctuation. 
Because the first two timings are
the ones that determine the curvature in the $O-C$ residuals, 
one may question whether the curvature is real. Future
observations may prove that  the orbital period
of \object{BH Lyn} increases, however, our opinion 
is that only two timings determined from patrol plates do not provide 
enough evidence for this.
We therefore determined an updated linear ephemeris 
without using the first two timings:
\begin{equation}
{\rm HJD_{min}}=2447180.33600(28)+0\fd155875577(14)E. \label{ephe}
\end{equation}
This ephemeris is shown by the solid line in Fig.\,\ref{oc}.
It is very similar to the ephemeris of Hoard \& Szkody (\cite{hs});
 the orbital period is only slightly larger 
and  the reference times differ by $\leq1$ min.

\begin{figure}[t]
\centering
\includegraphics*[width=8.8cm]{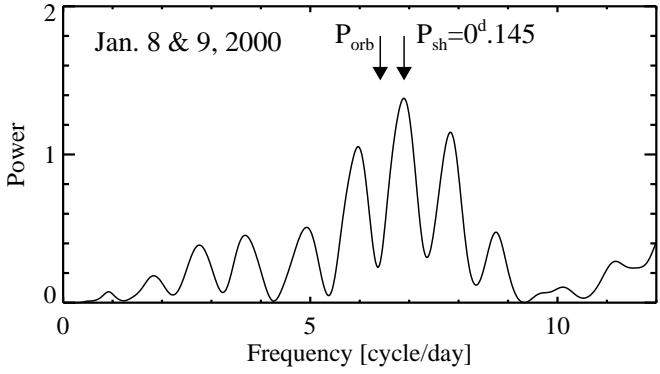} 
\caption{Periodogram of the January 2000 data. The negative superhump 
and the orbital periods are indicated.}
\label{per}
\end{figure}

The light curves show prominent humps whose maxima occur at different
orbital phases in the different runs (Fig.\,\ref{data}). 
We interpret this as an indication of the presence of
superhumps. Because our data are sparse, they are clearly not 
enough for an in-depth study of superhumps in
\object{BH Lyn}. After removing the data during eclipses, 
we computed the 
Lomb-Scargle periodogram (Scargle \cite{scar})
of the two January 2000 series only (Fig.\,\ref{per}).
 The strongest peak around the expected
frequency of the superhumps corresponds to a period of
 $\sim$0\fd1450\,$\pm0.0065$, which is 
close to the negative superhumps period 0\fd1490\,$\pm0.0011$ reported by Patterson
(\cite{patt99}). The least-squares fit gives the semi-amplitude of the signal of
0.084\,$\pm0.005$ mag.

\begin{figure}
\centering
\includegraphics*[width=8.8cm]{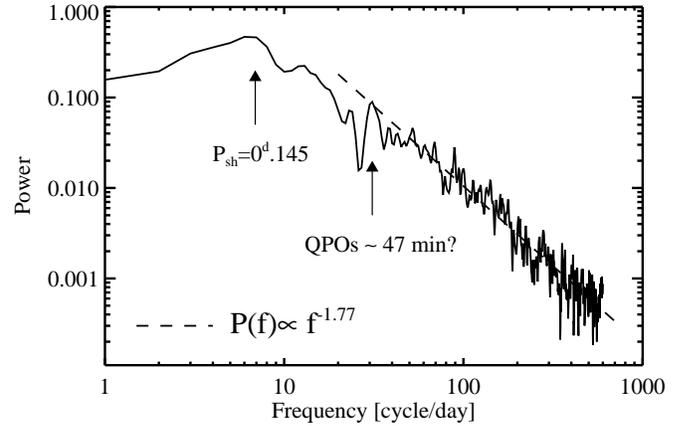}   
\caption{The mean power spectrum of \object{BH Lyn} light curves.}  
\label{slope}
\end{figure}

We have also searched all runs for periodic variations on the minute time-scale.
The power spectra show many peaks with frequencies below 
$\sim150\ {\rm cycle\,day}^{-1}$, but 
the attempts to fit the runs with periods corresponding
to any of the peaks in the periodograms were not satisfactory. 
Thus, most probably no coherent oscillations are present. 
The individual power 
spectra show a typical red noise shape characterized by a power-law 
decrease of the power with frequency $P(f)=f^\gamma$. 
The mean power spectrum of \object{BH Lyn} has  power-law index 
$\gamma=-1.77$. Because  the red noise 
processes have strong low-frequency variability, 
it is most likely that the peaks in the periodograms are due to the red noise. 
Nevertheless,
the peak at $\sim32\ {\rm cycles\,day}^{-1}$ is present in most 
periodograms, and it is also noticeable in the mean power spectrum 
(Fig.\,\ref{slope}). This might indicate the presence of quasi-periodic 
oscillations like the ones discussed by Patterson et al. (\cite{patt02}),
however, a study based on more data is needed to confirm
this. 
 
 The red noise in the power spectra of CVs is a result of flickering 
 (Bruch \cite{bruch}).
\object{BH Lyn} light curves show strong flickering activity; flickering peaks
with typical durations of  5--10 min and amplitudes reaching $\sim0.2$ mag can
be recognized in Fig.\,\ref{data}. 
The mean standard deviation in the light curves after the
low-frequency signals have been subtracted is $\sim0.06$ mag.
This value is consistent with the standard deviation found in the light
curves of the NLs \object{TT Ari}, \object{MV Lyr} and \object{PX And} 
(Kraicheva et al. \cite{krtt2},b; Stanishev et al. \cite{px}).

\begin{figure}[t]
\centering
\includegraphics*[width=8.8cm]{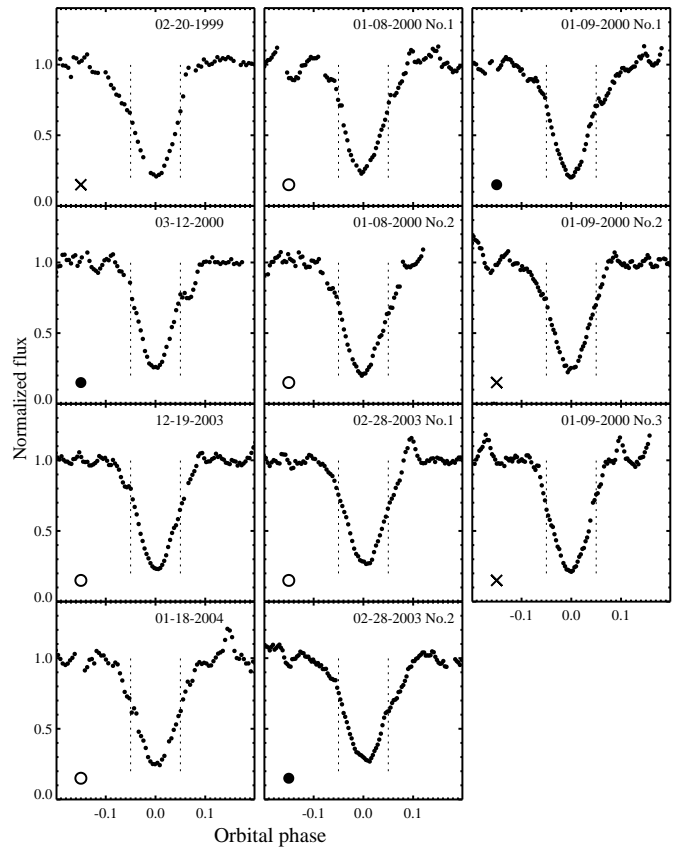}   
\caption{Normalized eclipses of \object{BH Lyn}. The dashed lines are guide to
the eye to see the difference of the eclipse profile easier. The symbols used
for the eclipses in Fig.\,\ref{ecl1s} are shown in the lower left corners.}
\label{ecl}
\end{figure}

\begin{figure}[t]
\centering
\includegraphics*[width=8.8cm]{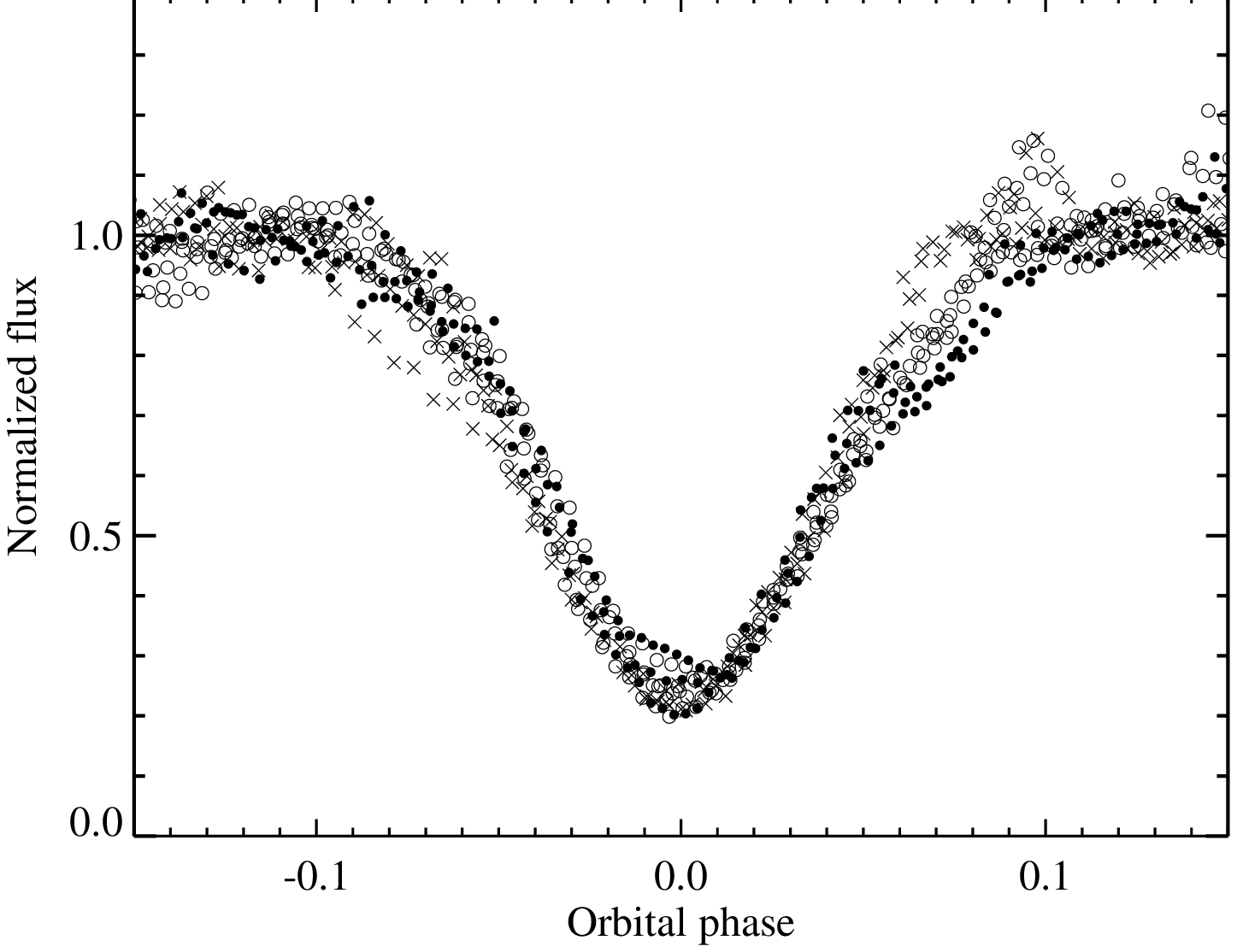}  
\caption{The three eclipse groups plotted together with different symbols.}
\label{ecl1s}
\end{figure}

The depth of the eclipses in \object{BH Lyn} during our observations 
is $\sim1.5$ mag, and their average  
{\it full-width at half-flux} is  $0.0683(\pm0.0054)P_{orb}$.
The out-of-eclipse magnitudes were fitted with
low-order polynomial functions to account for  brightness
variations that are not due to the eclipse, but most probably arise
from the superhumps. The eclipses were normalized to the fits and are
shown in Fig.\,\ref{ecl}. As can be seen, there is a substantial variability 
of the eclipse shape, even during a single night. The variations  
are most notable in the upper half of the eclipse profiles. Half of the eclipses 
appear to be fairly symmetric, while the rest are clearly asymmetric.
 More interestingly, though, the egress of the eclipses on Mar. 3, 2000 and 
No.1 on Jan. 9, 2000, and possibly the ingress of some other 
 eclipses, are not monotonic. To highlight the differences, in Fig.\,\ref{ecl1s}
we show all the eclipses together. Except for the single eclipse
in 1999, the ingress of all eclipses are very similar. The egress of the eclipse are however
very different, and the eclipses could be split into three sequences.
In Fig.\,\ref{ecl1s}, each of these groups is plotted with a different symbol.

\section{Discussion}

Because of the large variability of the eclipse profiles in \object{BH Lyn},
we are reluctant to attempt eclipse mapping or to try to estimate the system
parameters from the eclipse width. Clearly, such analyzes could give false results.
The rather rapid changes in the eclipse profiles, even during a single night,
could be explained by temporal variations of the AD 
surface brightness distribution. Large flickering peaks can 
be seen before or after some of the eclipses (Fig.\,\ref{ecl}). 
If such a peak occurs during an eclipse, it could
alter its shape, even to cause the eclipse not to be monotonic.
Another explanation could be that the amount of 
overflowing gas varies, and as a consequence the intensity of
the two hot spots could also change, causing variations in the eclipse profile. 

Variations of the area of the eclipsing body with time will also cause
variations of the eclipses. Given the time scale of the observed 
changes, the secondary is ruled out. On the other hand, the 
SW Sex stars most likely possess very complex accretion structures, and  
it may be that the AD is self-occulting. Self-occultation
seems to be the most reasonable explanation of the UV observations of 
another SW Sex star, \object{DW UMa} (Knigge et al. \cite{dwuma}), hence giving 
support for this in \object{BH Lyn}. Variations of the effective area of 
the occulting parts may cause the observed eclipse profile changes.

The presence of negative superhumps in eclipsing SW Sex stars is 
very interesting. The origin of negative superhumps is still 
a puzzle, but they are believed to be caused by a retrograde precession of 
an accretion disc (AD) that is tilted with respect to the orbital plane 
 (Bonnet-Bidaud et al. \cite{tilt}). 
If negative superhumps do arise from the precession of tilted ADs, then the 
accretion stream overflow would easily  occur (Patterson et al.
\cite{patt97}). 
Therefore,  the SW Sex and negative superhumps phenomena 
should have the same origin. 
Due to the presence 
of precessing tilted AD,  the amount of gas in the 
overflowing stream will be modulated on the negative superhump period. Hence,
the intensity of the second hot spot will change and may produce superhumps
(Patterson et al. \cite{patt97}; Stanishev et al. \cite{px}).
This scenario can be observationally tested. In this model,
the negative superhumps should manifest themselves in spectra in two ways:
1) the intensity of the high-velocity emission components in spectra, which are
thought to arise from the second spot, should
be modulated with the superhumps period; 2) 
since the orientation of the tilted disc with respect to the 
observer will change over the precession cycle, at certain precession 
phases, the SW Sex signatures should disappear. 
 To test these predictions, time-resolved high
signal-to-noise spectrophotometry over several consecutive nights is needed,
since the precession periods are of
the order of a few days. We encourage such studies.

\begin{acknowledgements}
The work was partially supported by NFSR under project No.~715/97.
\end{acknowledgements}

\end{document}